\documentclass[aps,pra,showpacs,twocolumn]{revtex4}
\usepackage{graphicx}
\usepackage{bm}
\usepackage{amssymb}
\usepackage{times}

\newcommand{\trace}{\mathop{\rm Tr}\nolimits}

\newcommand{\diag}{\mathop{\rm Diag}\nolimits}

\newcommand{\conv}{\mathop{\rm conv}\nolimits}

\newcommand{\bra}[1]{\langle#1|}
\newcommand{\ket}[1]{|#1\rangle}

\newcommand{\twomat}[4]{\left(\begin{array}{cc}#1&#2\\#3&#4\end{array}\right)}

\newcommand{\schatten}[2]{\left|\left|\,{#2}\,\right|\right|_{#1}}

\newcommand{\cH}{{\mathcal H}} % \cal not known
\newcommand{\cM}{{\mathcal M}} % \cal not known
\newcommand{\cN}{{\mathcal N}} % \cal not known
\newcommand{\cT}{{\mathcal T}} % \cal not known

\newcommand{\qed}{\hfill$\square$\par\vskip12pt}

\newcommand{\G}{Random Unitary} % name for mixed unitaries

\newcommand{\C}{{\mathbb{C}}}

\DeclareRobustCommand\openone{\leavevmode\hbox{\small1\normalsize\kern-.33em1}}
\newcommand{\id}{\mathrm{\openone}}

\newcommand{\be}{\begin{equation}}
\newcommand{\ee}{\end{equation}}
\newcommand{\bea}{\begin{eqnarray}}
\newcommand{\eea}{\end{eqnarray}}
\newcommand{\beas}{\begin{eqnarray*}}
\newcommand{\eeas}{\end{eqnarray*}}

\newtheorem{theorem}{Theorem}
\newtheorem{lemma}{Lemma}
\newtheorem{conjecture}{Conjecture}

\newtheorem{property}{Property}

\newcount\minute
\newcount\hour
\def\currenttime{%
    \minute\time
    \hour\minute
    \divide\hour60
    \the\hour:\multiply\hour60\advance\minute-\hour\the\minute}

\begin{document}
\title{On Random Unitary Channels}
\author{Koenraad M.R.~Audenaert}
\email{k.audenaert@imperial.ac.uk}
\affiliation{Institute for Mathematical Sciences, Imperial College London,
53 Prince's Gate, London SW7 2PG, UK}
\affiliation{Dept. of Mathematics, Royal Holloway, University of
London, Egham, Surrey TW20 0EX, UK}
\author{Stefan Scheel}
\email{s.scheel@imperial.ac.uk}
\affiliation{Quantum Optics and Laser Science, Blackett Laboratory,
Imperial College London, Prince Consort Road, London SW7 2AZ, UK}

\begin{abstract}
In this article we provide necessary and sufficient conditions for a
completely positive trace-preserving (CPT) map to be decomposable into
a convex combination of unitary maps. Additionally, we set out to define
a proper distance measure between a given CPT map and the set of random
unitary maps, and methods for calculating it. In this way one could
determine whether non-classical error mechanisms such as spontaneous
decay or photon loss dominate over classical uncertainties, for
example in a phase parameter.
The present paper is a step towards achieving this goal.
\end{abstract}

\date{\today, \currenttime}

\maketitle

%%%%%%%%%%%%%%%%%%%%%%%%%%%%%%%%%%%%%%%%%%%%%%%%%%%%%%%%%%%%%%%%%%%%%%
\section{Introduction}
In this paper we answer two questions about quantum maps. The first
question is, given any completely positive trace-preserving (CPT) map,
how can one determine whether this map can be decomposed as a convex
combination of unitary maps? In more formal terminology we ask for
necessary and sufficient conditions such that a CPT map $\Phi$ can be
written as
\be
\rho\mapsto\Phi(\rho) = \sum_i p_i\,U_i \rho U_i^\dagger,
\ee
where the scalars $p_i$ form a probability distribution (i.e.\ they
are non-negative and add up to 1) and where the $U_i$ are unitaries.
Preferably, the method for doing so should be constructive and should
as a bonus supply the $p_i$ and $U_i$.
CPT maps obeying this condition are called \textit{\G}\ maps. The
second, and very much related, question we answer here is about
finding a proper distance measure between a CPT map and the set of \G\
maps, and methods for calculating it, be it numerical or (preferably)
analytical ones.

The physical motivation behind these questions is the desire to
distinguish various error mechanisms afflicting the preparation and
processing of quantum states. For example, errors occur in the
realisation of quantum gates in quantum information processing.
If the only error mechanism occurring is a classical uncertainty, for
instance in a phase parameter, then the resulting ``gate'' will not be
described by a particular unitary, but rather by a mixture of such
unitaries; the mathematical description of such a mixture is
effectively a \G\ map. If on the other hand, other mechanism can
occur, such as spontaneous decay or photon loss, then the resulting
gate can no longer be described by such a \G\ map. In a sense, the
distance between this particular map and the set of \G\ maps
determines and quantifies the presence of these non-classical error
mechanisms.

The paper is structured as follows. In Section \ref{sec:prelim}, we
present notations and definitions for a number of basic concepts that
will be needed in the rest of the paper. Section \ref{sec:det} is
devoted to the determination of whether a unital CPT map is a random
unitary map. The techniques introduced in this Section are then
generalised in Section \ref{sec:dist} to construct a genuine distance
measure $D$ between a map and the set of \G\ maps. A number
of properties of $D$ are subsequently derived. Finally, in Section
\ref{sec:eoa}, a connection is made to the entanglement of assistance
of bipartite states.

%%%%%%%%%%%%%%%%%%%%%%%%%%%%%%%%%%%%%%%%%%%%%%%%%%%%%%%%%%%%%%%%%%%%%%%
\section{Preliminaries}\label{sec:prelim}

It is quite obvious that any \G\ map should, apart from being
trace-preserving, also be unital, meaning that $\Phi(\id)=\id$. Maps
like this, trace-preserving and unital ones, are also called
\textit{doubly stochastic}. This necessary condition of double
stochasticity implies that the input and output dimensions of the map
should be identical. We shall henceforth assume that $\Phi$ is a
unital CPT map taking states on a $d$-dimensional Hilbert space $\cH$
to states on that same space.

%%% Landau & Streater
At this point it is of course very tempting to check whether the
condition of double stochasticity may even be necessary. In
fact, for qubit CP maps ($d=2$) this is the case. By definition, the
set of doubly stochastic CP maps is convex. In \cite{streater}, Landau
and Streater proved that for $d=2$ the extremal points of this convex
set are precisely the unitary maps. This is just a reformulation of
the statement that the set of \G\ qubit-maps is precisely the set of
doubly stochastic CP qubit-maps.

For higher dimensions this is no longer true, as was first shown by
Tregub and by Kummerer and Maassen
\cite{tregub,maassen,streater}. That is, for higher $d$ there are
extremal doubly stochastic CP maps that are not unitary. An example of
such a map in odd dimensions (taken from Ref.~\cite{streater}) is the
following:
$$
\rho\mapsto \Phi(\rho) = \frac{1}{j(j+1)} \sum_{k=1}^3 J_k \rho J_k,
$$
where $d=2j+1$ and the $J_k$ are the three well-known generators of
$SU(2)$ in its $d$-dimensional irreducible representation \cite{greiner}.
Consequently, the set of all convex combinations of unitary maps is
a proper subset of the set of all doubly stochastic CP maps.

%%%%%%%%%%%%%%%%%
\subsection{Jamio{\l}kowski Isomorphism}

To proceed, we will next exploit the Jamio{\l}kowski isomorphism between
CP maps and states. For the purposes of this paper it is not strictly
necessary to do so, but it has the benefit of widening the perspective.
The Jamio{\l}kowski isomorphism assigns to each CP map from
$\cH_{\text{in}}$ to $\cH_{\text{out}}$ a (not necessarily normalised)
state on $\cH_{\text{out}}\otimes\cH_{\text{in}}$; this state will be
called here the \textit{state representative} of the CP map, and will
be denoted by $\Phi$ as well. The context will make clear whether the
map is meant or its state representative. To avoid confusion, we will
always assume that the state representative is normalised (i.e.\ it
has unit trace). The density matrix representing the state
representative is sometimes called the Choi matrix, after M.-D. Choi
who proved that under the Jamio{\l}kowski isomorphism CP maps are mapped
to positive semidefinite matrices \cite{choi}, which is not true for
any non-CP map.

In our case, the state representative is a $d\times d$ state.
To explain how the assignment is done, we first introduce the symbol
$\ket{I}$ for the maximally entangled state vector
\be
\ket{I} := \frac{1}{\sqrt{d}} \sum_{i=1}^d \ket{i,i}.
\ee
Then the following formula defines the Jamio{\l}kowski isomorphism:
\be
\Phi = (\Phi\otimes\id)(\ket{I}\bra{I}),
\ee
where $\id$ stands for the identity map; so the map $\Phi$ operates on
the first copy of $\cH$, while the second copy is left untouched (is
``operated'' upon by the identity map).

As our map is trace-preserving, its state representative is
automatically normalised. Moreover, the reduction of the state
representative obtained by tracing out the first copy of $\cH$ yields
the maximally mixed state $\id/d$. Furthermore, unitality of the map
shows itself through the fact that the second reduction (tracing out
the second copy of $\cH$) is also the maximally mixed state.

Applying this game to the question under consideration yields that a
unital CPT map is a \G\ map iff its state representative is of the form
\be
\Phi = \sum_i p_i (U_i\otimes\id) \, \ket{I}\bra{I} \,
(U_i^\dagger\otimes\id).
\ee
This actually says that the state representative should be a mixture
of maximally entangled (ME) pure states. Indeed, any such pure state
can be obtained from the ``vintage'' ME state $\ket{I}$ by applying a
local unitary on either party. The set of such mixtures forms, by its
very definition, a convex set with the ME pure states as extremal
points. We will denote this set by the symbol $\cM$.

Furthermore, we will denote the set of states with maximally mixed
reductions by $\cN$:
\be
\cN := \{\rho: \trace_1 \rho = \trace_2\rho=\id/d\}.
\ee
Obviously, we have $\cM\subset\cN$. We will see below that for
$2\times2$ states, $\cM$ is actually equal to $\cN$, while this is no
longer the case for higher-dimensional states.

Our questions have therefore been reduced to determining whether a
state is a mixture of ME pure states or, if that is not possible, how
far the state is from the convex set $\cM$. What almost immediately
comes to mind is the resemblance between these questions and the
notorious questions of determining separability and entanglement,
where the convex set under consideration is the convex hull of all
pure \textit{product} states. To wit, what we do in this paper is
nothing but adapting \textit{mutatis mutandis} various methods from
entanglement theory to the problem at hand.

In the next few paragraphs we introduce some more notations that will
simplify the rest of the presentation.

%%%%%%%%%%%%%%%%%%%%%%%%%%
\subsection{Matrixification}

An important linear operation one can perform on bipartite state
vectors is the ``\textit{matrixification}'' operation \cite{ka01}. For
a given bipartite $d\times d$ state vector $\ket{\psi}$, we define its
matrixification as a $d\times d$ matrix, denoted by $\tilde{\psi}$,
such that the following holds:
\be
\ket{\psi} = (\tilde{\psi}\otimes\id)\ket{I}.
\ee
Put less abstractly, the matrix $\tilde{\psi}$ consists of all the
vector entries of $\psi$ placed in a matrix frame. That is, if
$\ket{\psi}$ is given by $\ket{\psi} = \sum_{i,j} a_{ij}\ket{i,j}$
then $\tilde{\psi}$ is just the matrix
$a=\sum_{i,j} a_{ij}\ket{i}\bra{j}$.

In passing, we remark that the Schmidt decomposition of a bipartite
pure state is nothing but the singular value decomposition (SVD) of
the matrixified state vector, and the Schmidt coefficients are just
the singular values.

%%% dual CP maps, TP-ness and unitality
For every linear map $\Phi$, one defines the dual map $\Phi'$ as that
map for which $\trace[A\Phi(B)] = \trace[\Phi'(A) B]$ for all $A,B$.
If $\Phi$ is CP, then so is $\Phi'$. Furthermore, if $\Phi$ is
trace-preserving then $\Phi'$ is unital, and if $\Phi$ is unital then
$\Phi'$ is trace-preserving.

%%% Bloch vectors
\subsection{Bloch Vector Formalism\label{sec:bloch}}

We will also have the opportunity to use the well-known Bloch vector
formalism for representing states. The Bloch vector formalism is based
on the observation that the density matrices on $\cH$ are themselves
embedded in a Hilbert space, a space of dimension $d^2$. Because
density matrices are Hermitian, this Hilbert space is real. The inner
product in that space is given by the functional
$\langle\rho,\sigma\rangle:=\trace[\rho\sigma]$.
The Bloch vector of a density matrix is just the vector representing
that matrix in this Hilbert space, the entries of which of course
depend on the choice of basis.

The standard choice of basis elements are the Pauli matrices
$\sigma_i$ (for $d=2$) and generalisations thereof to higher
dimensions; e.g., in $d=3$ one has the Gell-Mann matrices. We use the
notation $\tau_i$ for these generalisations.
They can be grouped into $x$, $y$ and $z$ groups \cite{greiner}:
\beas
\tau_{x;kl}&=&\ket{k}\bra{l}+\ket{l}\bra{k}, \quad 1\le k<l\le d\\
\tau_{y;kl}&=&i(\ket{k}\bra{l}-\ket{l}\bra{k}), \quad 1\le k<l\le d\\
\tau_{z;k}&=&\sqrt{\frac{2}{k^2+k}}\diag(1^{\times k},-k,0^{\times d-k-1}), \\
&& \qquad\qquad 1\le k\le d-1.
\eeas
Here, the notation $a^{\times k}$ stands for $k$ entries equal to
$a$. It is standard to set $\tau_0=\sqrt{2/d}\id$. The inner products
between these matrices are given by
$\trace[\tau_i\tau_j]=2\delta_{i,j}$ (the factor 2 is conventional).

The entries of the Bloch vector in this basis are then given by
$\vec{\rho}:=(\rho_0,\rho_1,\ldots,\rho_{d^2-1})$
where $\rho_i:=\trace[\rho\tau_i]/2$; conversely,
$\rho=\sum_i \rho_i\tau_i$.
One easily checks the relation
$\trace[\rho\sigma] = 2\langle\vec{\rho},\vec{\sigma}\rangle$.

For normalised states, $\rho_0=1/\sqrt{2d}$, which is a constant. It
is therefore meaningful to employ the reduced Bloch vector instead,
which is $\tilde{\vec{\rho}}:=(\rho_1,\ldots,\rho_{d^2-1})$.
For normalised states, we have $\trace[\rho\sigma]$
$\!= 2\langle\tilde{\vec{\rho}},\tilde{\vec{\sigma}}\rangle+1/d$.

When the state $\rho$ is subjected to a unitary conjugation,
$\rho\mapsto U\rho U^\dagger$ ($UU^\dagger=\id$), the corresponding
Bloch vector will be rotated according to a certain orthogonal matrix
$O$ ($OO^T=\id$). Let $\rho=\sum_i \rho_i\tau_i$, then
$\rho'=U\rho U^\dagger = \sum_i \rho_i U\tau_i U^\dagger$  and
\beas
\rho'_j &=& \trace[\rho' \tau_j]/2 \\
&=& \sum_i \rho_i \trace[U\tau_i U^\dagger\tau_j]/2 \\
&=& \sum_i O_{ji}\rho_i,
\eeas
or $\vec{\rho'} = O\vec{\rho}$, where
$O_{ji}=\trace[U\tau_i U^\dagger\tau_j]/2$.
This defines a real matrix $O$ (because Bloch vectors are real), and
we need to show that $O$ is orthogonal. That is easily done as follows:
\beas
(OO^T)_{ij} &=& \sum_k O_{ik} O_{jk} \\
&=& \sum_k \trace[U\tau_k U^\dagger\tau_i] \,\trace[U\tau_k
U^\dagger\tau_j]/4 \\ 
&=& \trace[U \left(\sum_k\tau_k\trace[U\tau_k
U^\dagger\tau_j]/2\right) U^\dagger\tau_i]/2 \\ 
&=& \trace[U \left(U^\dagger\tau_j U\right) U^\dagger\tau_i]/2 \\
&=& \trace[\tau_j \tau_i]/2 = \delta_{ij},
\eeas
so that, indeed, $OO^T=\id$. Furthermore, as the unitary conjugation
leaves $\tau_0$ invariant (it is a multiple of the identity), $O$
decomposes as $[1]\oplus \tilde{O}$, where $\tilde{O}$ operates on the
reduced Bloch vector only.

Positivity of the density matrix translates to certain conditions on
the Bloch vector. It implies that the purity $\trace[\rho^2]$ of
states lies between $1/d$ and 1, so that the length of the Bloch
vector is bounded from above by $1/\sqrt{2}$. Likewise, the maximal
length of the reduced Bloch vector is $\sqrt{(d-1)/2d}$. States whose
reduced Bloch vector is this long are automatically pure. This means
that the reduced Bloch vectors of states all lie in a ball of radius
$\sqrt{(d-1)/2d}$, the so-called Bloch ball.

For $d=2$, all points in the Bloch ball turn out to correspond to
states. For higher $d$ this is no longer the case, and the set of
Bloch vectors of states is more like a ``dimpled'' ball, the
conditions on positivity leading to slices being cut away of the
original ball. It is easy to see that this has to be so: the vector of
diagonal entries of a density matrix is a probability vector and thus
lies on a simplex, which means that there must be a linear projection
under which the set of all Bloch vectors turns into that simplex. This
is impossible for a ball, unless that simplex is 1-dimensional, as it
is in the $d=2$ case.

%%%%%%%%%%%%%%%%%%%%%%%%%%%%%%%%%%%%%%%%%%%%%%%%%%%%%%%%%%%%%%%%%%%%%
\section{Determining whether a unital CPT map is a \G\ map}
\label{sec:det}

To answer the question of whether a unital CPT map is a \G\ map, we
freely borrow the methods described in \cite{ka01} and adapt it to the
problem at hand.

%%%%%%%%%%%%%%%%%%%%%%%%%%%%%%%%%%%%%%%%%%%%%%%%%%%%
\subsection{Condition for \G-ness: pure case}

The first step in answering the question is to find a quadratic
criterion for judging whether a pure state is maximally entangled
(ME). This is quite simple, and the answer is that a state vector
$\ket{\psi}$ is ME iff $\sqrt{d}\tilde{\psi}$ is unitary. Indeed, we
noted that ME states are characterised as
$\ket{\psi}=(U\otimes\id)\ket{I}$ for some unitary $U$,
so we trivially get $\tilde{\psi}=U$.

Thus, one must have
\be
\tilde{\psi}\tilde{\psi}^\dagger = \id/d.
\ee
The left-hand side is easily seen to be the reduction of the state
$\ket{\psi}\bra{\psi}$ to the first subsystem; thus the condition can
also be phrased as
\be\label{eq:condpure2}
\trace_2 \ket{\psi}\bra{\psi} = \id/d.
\ee
It is immediate from the purity of the state that the reduction to the
other subsystem will automatically be $\id/d$ as well. For normalised
$\psi$, this leaves us with a system of $d(d+1)/2-1$ quadratic equations.

In terms of CPT maps, pure states are the state representatives of
rank-1 CP maps, i.e. maps with a single element in their Kraus
decomposition. The above discussion then leads to the rather obvious
fact that rank-1 CP maps are \G\ maps if and only if they are unital
and trace preserving.

For higher-rank CP maps, this is no longer the case.
Nevertheless, the condition Eq.~(\ref{eq:condpure2}) will play a
central role for such maps as well.
It will turn out to be convenient to rephrase that condition in terms
of the Bloch vector of the reduced state.
One easily sees that it is equivalent to the condition that the
reduced state has the
zero-vector as reduced Bloch vector; this also holds when the state is
non-normalised.
Thus, the condition is given by the system of equalities
\be\label{eq:condpure3}
\trace[\tau_i \trace_2 \ket{\psi}\bra{\psi} ]=0,\text{ for } 1\le i\le
d^2-1, 
\ee
or
\be\label{eq:condpure4}
\bra{\psi}\tau_i \otimes \id\ket{\psi} = 0,\text{ for } 1\le i\le d^2-1.
\ee

%%%%%%%%%%%%%%%%%%%%%%%%%%%%%%%%%%%%%%%%%%%%%%%%%%%%%%%%%%%%
\subsection{Decompositions of Mixed States}

Now we must use the pure-state criterion to see whether a mixed state
can be decomposed as a mixture of ME pure states.
This brings us to the second step of the method: determine all
possible convex decompositions of a state. The answer to that is
well-known indeed and has been discovered many times. Let the
eigenvalue decomposition (EVD) of the state $\rho$ be given by
$\rho=U\Lambda U^\dagger$, where $U$ is the unitary whose columns are
the eigenvectors $u_i$, and $\Lambda$ is a diagonal matrix, the
diagonal elements of which are the eigenvalues $\lambda_i$; these are
non-negative and add up to 1. Then any other convex decomposition of
$\rho$ is given by $\rho=WMW^\dagger$, where $W$ is no longer unitary
and need not even be square; its columns are the (normalised) vectors
of which the decomposition is built. The matrix $M$, though, is still
diagonal with non-negative diagonal elements adding up to 1; these are
the convex weights of the decomposition.
Alternatively, these weights can be absorbed in their corresponding
vectors so that the norm squared of each vector then equals its
weight. This gives $\rho=ZZ^\dagger$, with $Z=WM^{1/2}$. In this way,
each convex decomposition of $\rho$ is related to a certain
``square-root'' of $\rho$.

Characterising all possible square roots of a matrix is a simple
problem in matrix analysis. Starting from the eigenvalue square root,
$\rho^{1/2}=U\Lambda^{1/2}U^\dagger$, one generates all others by
right-multiplying it with a \textit{right-unitary} matrix $T$. A
right-unitary matrix $T$ is a non-necessarily square matrix for which
$TT^\dagger=\id$ holds (but not $T^\dagger T=\id$, unless $T$ is
square). Thus one has
\be\label{eq:ZT}
Z = \rho^{1/2}T.
\ee
The number of columns in $T$ is the number of vectors in the
decomposition, and is called the \textit{cardinality} of the
decomposition; this number should be at least as large as the rank of
$\rho$, i.e.\ the number of non-zero eigenvalues.

In quantum physics circles, the most recent and probably best-known
incarnation of this result is the famous Hughston-Jozsa-Wootters (HJW)
theorem \cite{hjw}. The earliest occurrence is actually in
Schr\"odinger's work on quantum steering \cite{schrodinger},
and it has been rediscovered many times by various physicists (see
Ref.~\cite{kirk} and references therein).

We now combine the HJW theorem with the quadratic characterisation of
pure ME states to obtain a method for determining \G-ness.

%%%%%%%%%%%%%%%%%%%%%%%%
\subsection{Criterion for \G-ness: mixed case\label{sec:Gmixed}}

Let us consider a particular convex decomposition of our map, or of
its state representative $\rho$, that is described by the matrix
$Z$. As $\rho$ is a $d\times d$ state, $Z$ has $d^2$ rows. The
cardinality $K$ of the decomposition, being the number of columns of
$Z$, cannot be smaller than the rank $R$ of $\rho$. We want to check
if all the vectors in this decomposition are ME state vectors. Thus we
have to apply the criterion (\ref{eq:condpure4}) to every column
vector of $Z$. This yields the system of equalities
\be
(Z^\dagger(\tau_i\otimes\id)Z)_{jj}=0,\text{ for } 1\le j\le K, 1\le
i\le d^2-1. 
\ee
Inserting Eq.\ (\ref{eq:ZT}) gives
\be
(T^\dagger \rho^{1/2}(\tau_i\otimes\id)\rho^{1/2} T)_{jj}=0,
\ee
for $1\le j\le K, 1\le i\le d^2-1$.
For succinctness we will introduce the $d^2-1$ matrices $A_i$ defined by
\be\label{eq:defA}
A_i:=\rho^{1/2}(\tau_i\otimes\id)\rho^{1/2}.
\ee

The problem is thus reduced to the following:

\medskip

\noindent
\textbf{Problem P:}\textit{
Find a scalar $K\ge d^2$ and a right-unitary $d^2\times K$ matrix $T$
that ``off-diagonalises'' the $d^2-1$ matrices $A_i$ (of dimension $d^2\times d^2$) simultaneously:
\be
\diag(T^\dagger A_i T)=0,\text{ for } 1\le i\le d^2-1.
\ee
}

Necessary conditions for such a $T$ to exist are that all $A_i$ should
be traceless. For the $A_i$ of Eq.~(\ref{eq:defA}) this means that
$\trace[\tau_i\trace_2[\rho]]=0$, i.e. $\trace_2[\rho]$ must be
proportional to the identity, which is a condition we have already
encountered.

In general, we do not know how to solve this problem analytically, and
we have to resort to numerical methods, just as in \cite{ka01}. This will
be described below in Section~\ref{sec:dist} on distance measures.

The off-diagonalisation problem can easily be solved for a single
matrix $A$. From the lemma below it follows that the necessary
condition $\trace[A]=0$ is also sufficient. Here, $K=d^2$ suffices,
and $T$ is a genuine unitary matrix.
\begin{lemma}
For a Hermitian $n\times n$ matrix $A$, a unitary $T$ exists such that 
$(T^\dagger AT)_{jj}=0$ for all $j$ if and only if $\trace[A]=0$.
\end{lemma}
\textit{Proof.}
Necessity is obvious as the trace is unitarily invariant.
To show sufficience, consider Schur's majorisation theorem \cite{HJI},
which says that for any Hermitian matrix $X$,
$$
\sum_{j=1}^k X_{jj} \le \sum_{j=1}^k \lambda_j^\downarrow(X),
$$
and equality holding for $k=n$.
Horn's Lemma \cite{HJI} adds to this that for any specified set of
diagonal entries and eigenvalues obeying this majorisation relation a
Hermitian matrix exists exhibiting those diagonal entries and
eigenvalues.

For fixed Hermitian $X$ the mapping $j\mapsto \lambda_j^\downarrow(X)$
is non-increasing by definition, so that the mapping
$k\mapsto \sum_{j=1}^k \lambda_j^\downarrow(X)$ is concave.
Now note that if $\trace[X]=0$ then $\sum_{j=1}^n
\lambda_j^\downarrow(X)=0$, while
$\sum_{j=1}^0 \lambda_j^\downarrow(X)=0$ trivially.
By the above concavity statement we then find that
$\sum_{j=1}^k \lambda_j^\downarrow(X)\ge0$ for all $k$.
As the zero vector $(0,\ldots,0)$ is majorised by any non-negative
vector, Horn's Lemma then implies the existence of a Hermitian matrix
with zero diagonal and any prescribed set of eigenvalues that add up
to 0.

Consider now the eigenvalues of $A$, which add up to 0 by
assumption. By the above, another matrix $X$ must exist exhibiting the
same eigenvalues, hence unitarily equivalent to $A$, and with zero
diagonal. Therefore, a unitary $T$ exists such that
$X_{jj}=(T^\dagger AT)_{jj}=0$ for all $j$.
\qed

For qubit maps there are three $A_i$ to cope with.
However, in the light of Landau and Streater's result that the qubit
\G\ maps are exactly the doubly stochastic maps, the condition
$\trace[A_i]=0$ should also be sufficient for $d=2$ and $A_i$ of the
form (\ref{eq:defA}). Indeed, the following Theorem holds, which
therefore supplies an alternative proof of Landau and Streater's
Theorem.
\begin{theorem}
Let $\rho$ be a $2\times2$ state with reductions $\rho_1=\rho_2=\id/2$;
let $\rho=U\Lambda U^\dagger$ be the eigenvalue decomposition of $\rho$;
let $W=U\Lambda^{1/2}$.
Let also $\sigma_i$ be the SU(2)-Pauli matrices. Then the diagonal
elements of $W^\dagger(\sigma_i\otimes\id)W$ are 0 for all $i$.
\end{theorem}
In other words, in this case we do not even have to search for the
matrix $T$.

\textit{Proof.}
The first thing to note is that the diagonal elements of
$W^\dagger(\sigma_i\otimes\id)W$ are 0 if and only if the diagonal
elements of $U^\dagger(\sigma_i\otimes\id)U$ are 0. As pure states in
$\cN$ are automatically in $\cM$, an equivalent statement of the
Theorem is that a 2-qubit state is in $\cN$ if and only if all its
eigenvectors are.

A $2\times2$ density matrix can be represented by a $2\times 2$ block
matrix:
$$
\rho = \twomat{B}{C}{C^\dagger}{D},
$$
where every block is a $2\times2$ matrix itself.
The conditions on $\rho$ then translate to
$$
B+D=\frac{1}{2}\id, \trace[B]=\trace[D]=1/2, \trace[C]=0.
$$
If we drop the normalisation condition on $\rho$, these conditions
relax to
$$
B+D=k\id, \trace[B]=\trace[D], \trace[C]=0,
$$
for some real number $k$.

We will now show that if $\rho$ satisfies these conditions, then its
square also does. Without loss of generality we can apply a local
unitary, so that $B$ can be diagonalised. Let us set
$$
B=\twomat{a}{0}{0}{b},
$$
with $a,b$ non-negative real numbers.
The conditions on $\rho$ then demand that $D$ is given by
$$
D=\twomat{b}{0}{0}{a}.
$$
Let us also put
$$
C=\twomat{z}{x}{y}{-z},
$$
where $x,y,z$ are complex numbers.

The square of $\rho$ is now given by
$$
\rho^2 = \twomat{B^2+CC^\dagger}{BC+CD}{(BC+CD)^\dagger}{C^\dagger C+D^2}.
$$
Let us now test the required conditions:
\beas
B^2+D^2+CC^\dagger+C^\dagger C&=&k'\id,\\
\trace[B^2+CC^\dagger]&=&\trace[D^2+C^\dagger C] ,\\
\trace[BC+CD]&=&0.
\eeas
Straightforward calculations reveal
$$
B^2+D^2 = (a^2+b^2)\id, CC^\dagger+C^\dagger C = (|x|^2+|y|^2+2|z|^2)\id,
$$
so that the first condition is satisfied.
Since $CC^\dagger$ and $C^\dagger C$ have the same trace, and
$\trace[B^2]=\trace[D^2]=a^2+b^2$, the second condition is also satisfied.
Finally, $\trace[BC+CD]=(a+b)z-(a+b)z=0$,
which shows that $\rho^2$ is of the required form.

We can now repeat the process of squaring and find that any $2^n$-th
power is of that same form. Let us now invoke the power method for
finding the dominating eigenvector $\psi$ of $\rho$: for $\rho$ with
non-degenerate spectra, its $m$-th power, after normalisation, tends
to $\ket{\psi}\bra{\psi}$ when $m$ tends to infinity.
By the above, we thus find that the projector on the dominating
eigenvector of $\rho$ is also in $\cN$.
Therefore, if we ``deflate'' $\rho$ by subtracting
$\lambda_1^\downarrow \ket{\psi}\bra{\psi}$ from it and renormalise,
we again obtain a state in $\cN$. Continuing in this way, we thus find
that every eigenvector of $\rho$ is in $\cN$, and hence in $\cM$.

For $\rho$ with degenerate spectra, continuity considerations lead to
the conclusion that one can always find vectors in its eigenspaces
that are in $\cM$.
\qed

%%%%%%%%%%%%%%%%%%%%%%%%%%%%%%%%%%%%%%%%%%%%%%
\subsection{Extremal CPT and UCPT maps}

The following theorem by Choi \cite{choi} characterises the extremal
CPT maps.
\begin{theorem}[Choi]
The CP map defined by
$\rho\mapsto\Phi(\rho)=\sum_{k=1}^R A_k^\dagger\rho A_k$ is extremal
within the set of CP maps with prescribed value of $\Phi'(\id)$ if and
only if the set
$$
\{A_k A_l^\dagger; k,l=1,\ldots,R\}
$$
of $R^2$ matrices is linearly independent.
\end{theorem}
For CPT maps, the requirement is $\Phi'(\id)=\id$.
As a simple consequence of this Theorem we note that the extremal
$d$-dimensional CPT maps have rank at most $d$. This is because at
most $d^2$ matrices of size $d\times d$ can be linearly
independent. In other words, the convex set of CPT maps on a
$d$-dimensional Hilbert space is the convex hull of the set of CPT
maps of rank at most $d$. Numerical experiments for $d$ up to 6 reveal
that in $d$ dimensions one can find $d$ matrices $A_k$ satisfying this
condition, so that the constraint $R\le d$ is saturated. Furthermore,
the condition is generically satisfied for a randomly generated set of
$d$ matrices $A_k$.

Let us now proceed to the study of extremal unital CPT maps. The
relevant generalisation of Choi's theorem is (\cite{streater}, Theorem
2):
\begin{theorem}[Landau-Streater]
The CP map defined by
$\rho\mapsto \Phi(\rho)=\sum_{k=1}^R A_k^\dagger\rho A_k$
is extremal within the set of CP maps with prescribed values of
$\Phi(\id)$ and $\Phi'(\id)$
if and only if the set of $R^2$ matrices (of size $2d\times 2d$)
$$
\{ A_k^\dagger A_l \oplus A_l A_k^\dagger; k,l=1,\ldots,R \}.
$$
is linearly independent.
\end{theorem}
This directly implies (\cite{streater}, Remark 3) that the extremal
maps have rank not higher than $\sqrt{2}d$. Again, numerical
experiments for $d$ up to 6 reveal that in $d$ dimensions one can find
$\lfloor\sqrt{2}d\rfloor$ matrices $A_k$ satisfying the condition of
the Theorem, so that the constraint $R\le \sqrt{2}d$ is saturated;
moreover, the condition is generically satisfied for a randomly
generated set of $\lfloor\sqrt{2}d\rfloor$ matrices $A_k$.

This remains particularly true for unital CPT maps, up to one
exception: for $d=2$ one cannot find more than 1 matrix $A_k$ obeying
the independence condition. This is in accordance with the statement
(\cite{streater}, Theorem 1) that for $d=2$ there are only rank-1
extremal maps. The existence of rank-2 extremal unital CPT maps is
prevented by the conditions for double stochasticity,
$A_1A_1^\dagger+A_2A_2^\dagger=\id$ and $A_1^\dagger A_1+A_2^\dagger A_2=\id$,
which imply that the singular value decompositions of $A_1$ and $A_2$
must be $A_1=U\Sigma_1 V^\dagger$ and $A_2=U\Sigma_2 V^\dagger$,
(with the same $U$ and $V$!) with $\Sigma_1^2+\Sigma_2^2=\id$.
Therefore, the set $\{ A_k^\dagger A_l \oplus A_l A_k^\dagger; k,l=1,2 \}$
is not independent. Indeed, two of its elements are equal:
\beas
A_1^\dagger A_2 \oplus A_2 A_1^\dagger &=&
V\Sigma_1\Sigma_2 V^\dagger \oplus U\Sigma_1\Sigma_2 U^\dagger \\
&=& A_2^\dagger A_1 \oplus A_1 A_2^\dagger.
\eeas

In this context, the following conjecture is of relevance \cite{RFWsite}.
The conjecture is supported by numerical evidence.
\begin{conjecture}[Audenaert-Ruskai]
Every $d\times d$ state $\rho$ can be written as an equal-weight average
of $d$ states $\rho_i$ (not necessarily different)
that are of rank at most $d$ and have partial traces $\trace_A\rho_i$
and $\trace_B\rho_i$
identical to those of $\rho$.
\end{conjecture}

\textbf{Remark.}
In numerical experiments one is confronted with the question of how to
generate random CPT maps, unital CP maps, and doubly stochastic CP
maps. The first two questions are readily solved: one generates a
random CP map, and then projects it onto the set of CPT maps or unital
CP maps, respectively. Here, the two respective projections are the
operations (performed at the level of the map's Choi matrix):
\beas
\Phi\mapsto (G_1 \otimes\id)\Phi(G_1 \otimes\id),&&
G_1=(\trace_2[\Phi])^{-1/2} \\ 
\Phi\mapsto (\id\otimes G_2)\Phi(\id \otimes G_2),&&
G_2=(\trace_1[\Phi])^{-1/2}. 
\eeas
Note that these projections preserve CP-ness and do not increase the
rank of $\Phi$.

The question of how to generate random unital CPT maps is slightly
harder, as one has to satisfy the two constraints of TP-ness and
unitality at once. Fortunately, this can also be done using a
projection method. The method, called ``projections on convex sets''
(POCS), consists of an iterative scheme whereby the two projections
$G_1$ and $G_2$ are alternatingly applied to an initial CP map. It
turns out that, due to the convexity of the two sets, this process
converges very quickly \cite{borwein} to a CP map in their
intersection, i.e.\ to a CP map that is both CPT and unital.

%%%%%%%%%%%%%%%%%%%%%%%%%%%%%%%%%%%%%%%%%%%%%%%%%%%%%%%%%%%%%%%%%%%%%%%%
\section{Distance Measures}\label{sec:dist}

While up to this point we have been looking at conditions under which
a unital CPT map is a \G\ map, we can modify our method slightly to
calculate a kind of distance between a given map and the set of \G\
maps. In general, one can define distances between a point and a set
as the minimal distance between that point and any point from the
set. Choosing different point-to-point distance measures thus
\textit{induces} different point-to-set distance measures. The exact
choice of distance measure may be guided by considerations of physical
relevance or just of mathematical convenience.
An example of a distance measure with a clear physical meaning and
relevance is the gate fidelity \cite{gilchrist}. However, the
Hilbert-Schmidt norm distance between the state representatives is
arguably the simplest one when it comes to actually performing the
minimisation.

To calculate the chosen point-to-set distance, the point-to-point
distance has to be minimised over all points of the set. The problem
we encounter in our situation is that the set is defined in terms of
its extremal points, of which there is an infinite number. The
minimisation has thus to be performed over all possible convex
combinations of an infinite number of points. By Caratheodory's
theorem \cite{rockafellar}, only a finite number of points can have a
non-zero contribution to the convex combination. If the set is
embedded in a $d$-dimensional real space, the maximal number of points
required is $d$. In the present case (maps on $\C^d$), we are dealing
with $d^2\times d^2$ PSD matrices, which are embeddable in a
$d^4$-dimensional real space. Therefore, we need at most $d^4$ points
to make up the convex combination. Nevertheless, the minimisation
consists of varying $d^4-1$ real convex weights and $d^2-1$ real
parameters (to make up an $SU(d)$-unitary) for each of the $d^4$
extremal points, hence of the order of $d^6$ parameters in total.

In the following we take a different approach, by modifying the
treatment from the previous Sections such that a quantity emerges that
is more easy to calculate than induced point-to-set distance measures
but also has an interpretation as a distance measure. To do so, we
take the vectors of diagonal elements $\diag(T^\dagger A_i T)$,
concatenate them into a single vector and then find the $T$ that
minimises a well-chosen norm of that vector. The minimal norm then
quantifies how far the given map is from the set of \G\ maps. At this
point, we cannot yet say that what we get in this way is a genuine
distance measure. What we do get already is that the given map is \G\
if and only if this minimal norm is 0.

First of all, we have to properly choose a vector norm.
We can, for example, choose a norm of the form
\beas
D_{p,q}(\rho,T) &:=& \left(\sum_j\left(\sum_i|
(T^\dagger A_i T)_{jj}|^p\right)^{q/p} \right)^{1/q} \\
&=& \schatten{q}{\left(\schatten{p}{((T^\dagger A_i T)_{jj})_i}\right)_j},
\eeas
and its minimal value
$$
D_{p,q}(\rho):=\min_T D_{p,q}(\rho,T),
$$
which is to serve as distance quantification between the given map and
the set of \G\ maps. In what follows, we will restrict our attention to
$D_{2,1}$, which we denote by $D$ without subscripts, because it has a
number of desirable mathematical properties. For instance, it can be
redefined as the convex hull of a simple function on pure states (see
Property \ref{prop:convhull} below).

Calculating $D$ now requires minimisation over $T$, which is a
right-unitary matrix of dimension $d^2\times K$ with $K$ at most
$d^4$; again there are of the order of $d^6$ parameters, but in this
case they are all contained in a single mathematical object, a
right-unitary matrix, rather than in several unitaries and a number of
convex weights. The minimisation therefore has a simpler mathematical
structure, which leads to simplifications at the level of actual
algorithms, but also concerning the derivation of its basic
properties. In \cite{ka01}, a modified conjugated gradient method is
described for minimising functionals over the manifold of
right-unitary matrices. This method is directly applicable to the
problem at hand, and we have implemented it in Matlab \cite{suppl}.

In the rest of this Section, we discuss various properties of $D$,
including lower and upper bounds that are easy to calculate.

\begin{property}\label{prop:1}
$D(\rho)$ is invariant under ``local'' unitaries, that is, unitaries
operating on input or output space separately.
\end{property}
\textit{Proof.}
When $\rho$ is subjected to local unitary rotations, $\rho\mapsto
W\rho W^\dagger$, where $W=U\otimes V$,
the $A_i$ matrices will transform according to
\beas
A_i &\mapsto& W\rho^{1/2}W^\dagger (\tau_i\otimes\id)W \rho^{1/2}W^\dagger \\
&=& W\rho^{1/2} (U^\dagger\tau_i U\otimes\id) \rho^{1/2}W^\dagger.
\eeas
The ``outer'' appearence of the unitary conjugation will of course be
absorbed in the unitary $T$ in $(T^\dagger A_i T)$ and plays no
further role for determining $D(\rho)$. The unitary conjugation on the
$\tau_i$ corresponds to replacing $\tau_i$ by $\sum_k O_{ik} \tau_k$
($1\le i\le d^2-1$), for some real orthogonal matrix $O$ (see Section
\ref{sec:bloch}). The entries $(T^\dagger A_i T)_{jj}$ are replaced
accordingly by $\sum_k O_{ik} (T^\dagger A_k T)_{jj}$, which amounts
to rotating each one of the vectors $((T^\dagger A_i T)_{jj})_i$ (for
all $j$). As rotations leave the length ($\ell_2$-norm) of a vector
unchanged, this shows that $D(\rho)$ is indeed invariant under local
unitaries.
\qed

\begin{property}\label{prop:2}
The value of $D$ for pure $d\times d$ states is given by
\bea
D(\ket{\psi}\bra{\psi}) &=& \sqrt{2(\trace[(\trace_B
\ket{\psi}\bra{\psi})^2]-1/d)} \\
&=& \sqrt{2}\,\, \|\trace_B \ket{\psi}\bra{\psi}-\id/d \|_2.\label{eq:aaa}
\eea
The maximal possible value is $\sqrt{2(1-1/d)}$, a value that is
achieved for pure product states. The minimal possible value is 0,
which is achieved for ME pure states.
\end{property}
Note that in the context of \G\ maps, we are not directly interested
in $D$ on all possible pure states. Nevertheless, it can be
calculated and, moreover, it will be useful in what follows.

Note also that (\ref{eq:aaa}) applied to mixed states in $\cN$ always
yields 0, irrespective of whether they are in $\cM$ or not.

\medskip

\textit{Proof.}
We already know that $T$ will be of no influence, because a pure state
can only be realised in one way. We can therefore put $T=\id$.

For $\rho=\ket{\psi}\bra{\psi}$, the matrices $A_i$ are given by
\beas
A_i &=& \ket{\psi}\bra{\psi}(\tau_i\otimes\id)\ket{\psi}\bra{\psi}.
\eeas
Then the diagonal entries of $A_i$ are given by
$$
(A_i)_{jj} = |\psi_j|^2 \,\, \bra{\psi}(\tau_i\otimes\id)\ket{\psi}.
$$
Thus, using $\sigma$ as a shorthand notation for
$\trace_B(\ket{\psi}\bra{\psi})$,
\beas
D(\ket{\psi}\bra{\psi}) &=& \sum_j |\psi_j|^2 \,\,
 \|(\bra{\psi}(\tau_i\otimes\id)\ket{\psi})_i  \|_2 \\
&=&  \|(\bra{\psi}(\tau_i\otimes\id)\ket{\psi})_i  \|_2 \\
&=&  \|(\trace[\tau_i \,\sigma])_i \|_2 \\
&=& 2  \| \tilde{\vec{\sigma}} \|_2 \\
&=& 2\sqrt{(\trace[\sigma^2]-1/d)/2} \\
&=& \sqrt{2(\trace[\sigma^2]-1/d)}.
\eeas
\qed

%%%%%%%%%%%%%%%%%%%%%%%%%%%%%%%%%%%%%%%%%%%%%%

\begin{property}\label{prop:5}
The maximal possible value of $D$ for a state in $\cN$ is
$\sqrt{2(d-1)/d}$.
\end{property}
An example of a state achieving this value is the state
$\ket{0}\bra{0}\otimes\id/d$, which is the state representative of the CP map
representing photon loss as well as spontaneous decay of excited atoms:
$\Phi(\cdot)=\trace[\cdot] \ket{0}\bra{0}$.

\medskip

\textit{Proof.}
For convenience, we consider the state $\ket{d}\bra{d}\otimes\id/d$ instead.
The corresponding $A_i$ are given by
\beas
A_i &=& (\ket{d}\bra{d}\otimes\id)\,(\tau_i\otimes\id)\,
(\ket{d}\bra{d}\otimes\id)/d
\\
&=&\frac{(\tau_i)_{dd}}{d}\,\ket{d}\bra{d}\otimes\id.
\eeas
The only $\tau_i$ with non-vanishing $(d,d)$-entry is
$\tau_{z;d-1}$. Its $(d,d)$-entry is given by
$-\sqrt{2(d-1)/d}$. Thus, we get
\beas
D(\rho,T) &=& \sqrt{2(d-1)/d} \,\, \sum_j |(T^\dagger
(\ket{d}\bra{d}\otimes\id/d)T)_{jj}| \\
&=& \sqrt{2(d-1)/d} \,\,\trace[T^\dagger (\ket{d}\bra{d}\otimes\id/d)T] \\
&=& \sqrt{2(d-1)/d} \,\,\trace[\ket{d}\bra{d}\otimes\id/d] \\
&=& \sqrt{2(d-1)/d}.
\eeas
As this value is independent of $T$, this is also the value of $D(\rho)$.

By Property \ref{prop:2}, this is the maximally achievable value of
$D$ throughout, and what we have just shown is that this value is
achievable for states in $\cN$.
\qed

%%%%%%%%%%%%%%%%%%%%%%%%%%%%%%%%%%%%%%%%%%%%%%

\begin{property}\label{prop:convhull}\label{prop:3}
The function $D$ is the convex hull of its restriction to pure states.
That is:
\bea
D(\rho) &=& \min_{p_j,\psi_j} \bigg\{\sum_j p_j
D(\ket{\psi_j}\bra{\psi_j}): \nonumber \\
&& \qquad\quad \sum_j p_j \ket{\psi_j}\bra{\psi_j}=\rho \bigg\}.
\eea
As a direct consequence, $D$ is a convex function.
\end{property}
\textit{Proof.}
We proceed in much the same way as we did in Section \ref{sec:Gmixed},
where we obtained a criterion for \G-ness in the mixed case. Again we
need to consider all possible ensembles realising the state $\rho$,
which we can do by varying over all right-unitaries $T$ in the
expression $Z=\rho^{1/2}T$ (where $\rho=ZZ^\dagger$). Recall that the
$j$-th column of $Z$ is then $Z_j:=\sqrt{p_j}\ket{\psi_j}$. By this
variation over $T$ we minimise the quantity
\beas
\lefteqn{\sum_j p_j D(\ket{\psi_j}\bra{\psi_j})} \\
&=& \sum_j p_j  \|(\bra{\psi_j}(\tau_i\otimes\id)\ket{\psi_j})_i  \|_2 \\
&=& \sum_j  \| ((Z_j)^\dagger (\tau_i\otimes\id) Z_j )_i  \|_2 \\
&=& \sum_j  \| ((Z^\dagger (\tau_i\otimes\id) Z)_{jj} )_i  \|_2 \\
&=& \sum_j  \| ((T^\dagger A_i T)_{jj} )_i  \|_2 \\
&=& D(\rho,T).
\eeas
The minimisation over $T$ appearing in the convex hull construction
thus, indeed, yields $D(\rho)$.
\qed

\textit{Remark.}
Since the convex hull construction has a dual, we can find an
expression of $D$ involving a maximisation. The convex hull $\conv(f)$
of a function $f$ can be expressed as the Legendre transform of the
Legendre transform of $f$ \cite{rockafellar}.
That is, $\conv(f)=f^{**}$, where the Legendre transform $f^*$ of $f$
is, in the quantum context, defined as
$$
f^*(X) = \max_\rho \trace[X\rho]-f(\rho),
$$
where the maximisation is over all states $\rho$, and the argument $X$
is a Hermitian operator. In particular, we have
\beas
D(\rho) &=& \max_X \trace[X\rho]-D^*(X) \\
D^*(X) &=& \max_\psi \langle\psi|X|\psi\rangle - D(\psi),
\eeas
where the last maximisation is over pure states because $D$ is the
convex hull of $D$ restricted to the pure states.

\medskip

\begin{property}\label{prop:6}
A lower bound on $D(\rho)$ can be given in terms of the 2-norm
distance of $\rho$ to $\cM$:
\be
D_2(\rho,\cM) := \min_\sigma \left\{  \|\rho-\sigma \|_2:
\sigma\in\cM \right\}.
\ee
Namely,
\be
D(\rho) \ge
\sqrt{\frac{4}{d}\left(\frac{1}{2-D_2(\rho,\cM)^2}-\frac{1}{2}\right)}.
\ee
\end{property}
\textit{Proof.}
An upper bound on any norm distance of $\rho$ to $\cM$ is obtained by
restricting $\sigma$ to the extremal points of $\cM$,
which are the ME pure states. Recall that these can be parameterised
as $\ket{\phi}=(U\otimes\id)\ket{I}$ for unitary $U$.

Le us confine attention to pure $\rho=\ket{\psi}\bra{\psi}$ first.
Thus we are looking now at the minimisation
$$
\min_U
|\ \|\,\,\ket{\psi}\bra{\psi}-(U\otimes\id)
\ket{I}\bra{I}(U\otimes\id)^\dagger\,\|\ |.
$$
Let us now consider the 2-norm distance $D_2$.
Taking its square yields
\beas
D_2(\psi)^2 &=& \min_U 2-2|\langle\psi|(U\otimes\id)\ket{I}|^2 \\
&=& 2-2(\max_U |\trace[\tilde{\psi}^\dagger U]|/\sqrt{d})^2 \\
&=& 2-2 \|\tilde{\psi} \|_1^2/d.
\eeas
Note that normalisation of $\psi$ amounts to $ \|\tilde{\psi} \|_2=1$.
On the other hand,
\beas
D(\psi)^2 &=& 2 \|\trace_B \ket{\psi}\bra{\psi} -\id/d \|_2^2 \\
&=& 2 \|\tilde{\psi}\tilde{\psi}^\dagger -\id/d \|_2^2 \\
&=& 2( \|\tilde{\psi} \|_4^4-1/d).
\eeas

A simple application of H\"older's inequality yields the following:
for $X$ such that $ \|X \|_2=1$, $ \|X \|_4\ge  \|X \|_1^{-1/2}$.
Equality is achieved for $X=\id_d/\sqrt{d}$.

Thus we obtain, for pure states,
\beas
D &=& \sqrt{2( \|\tilde{\psi} \|_4^4-1/d)} \\
&\ge& \sqrt{2( \|\tilde{\psi} \|_1^{-2}-1/d)} \\
&=& \sqrt{(4/d)(1/(2-D_2^2)-1/2)}.
\eeas
It is easily checked that the RHS is a convex, increasing function of
$D_2$. Since $D_2(\rho)$ is by definition a convex function of $\rho$,
it follows that the RHS is a convex function of $\rho$ too. Now, $D$
is the convex hull of the restriction of $D$ to pure states, which
means that the largest convex function that coincides with $D$ on pure
states is $D$ itself. Therefore, the above inequality on pure states
readily extends to mixed states.
\qed

For small $D_2$, this lower bound simplifies to
$D(\rho)\ge D_2(\rho)/\sqrt{d}$.
An important consequence is that $D$ has the desirable property of
being approximately linear for states very close to $\cM$, just like
the distance measures $D_1$ and $D_2$. This tells us that $D$ can
indeed be interpreted as a distance (and not, for example, a power of
a distance).

\begin{property}\label{prop:4}
A lower bound on $D(\rho)$ is given by
$D(\rho)\ge\max(\sqrt{2} \|\trace_A\rho-\id/d \|_2,
\sqrt{2} \|\trace_B\rho-\id/d \|_2)$.
\end{property}
\textit{Proof.}
This is a simple consequence of Property \ref{prop:convhull}.
By definition, the convex hull of a function $f$ that is defined on
the pure states is the largest convex function coinciding with $f$ on
the pure states. The two functions appearing in the maximum are both
convex functions that coincide with $D$ on the pure states, and must
therefore be smaller than or equal to $D$.
\qed

\bigskip

For 2-qubit states that have rank 2, and hence for 2-element qubit
maps, it turns out that equality holds so that we get a closed-form
analytic expression for $D$:
\begin{property}\label{prop:7}
For 2-qubit states $\rho$ of rank 2,
$$
D(\rho)=\sqrt{2}\max( \|\trace_A\rho-\id/2 \|_2,  \|\trace_B\rho-\id/2
\|_2).
$$ 
\end{property}

While we have not been able to prove this yet, numerical experiments
indicate that the expression also holds for 2-qubit states of
arbitrary rank. Furthermore, this would be a simple consequence of the
Audenaert-Ruskai conjecture for $d=2$, combined with convexity of $D$
and the statement for rank-2 states.

\medskip

\textit{Proof.}
Let $\rho$ be a rank 2 state, with eigenvalue decomposition
$\rho=p|\psi\rangle\langle\psi| +(1-p)|\phi\rangle\langle\phi|$. To
cover all its realising ensembles of cardinality 2, we have to
consider all $2\times 2$ unitaries
$T=\twomat{e^{i\chi}\cos\theta}{\sin\theta}{-\sin\theta}{e^{-i\chi}\cos\theta}$.
We then get
$\rho=|\psi_1\rangle\langle\psi_1|+|\psi_2\rangle\langle\psi_2|$, with
the non-normalised states
\beas
\psi_1 &=& e^{i\chi}\cos\theta\sqrt{p}\,\psi - \sin\theta\sqrt{1-p}\,\phi, \\
\psi_2 &=& \sin\theta\sqrt{p}\,\psi + e^{-i\chi}\cos\theta\sqrt{1-p}\,\phi.
\eeas
We then have to calculate
$$
D = \min_{\theta,\chi} \sqrt{2}\sum_{i=1,2} \schatten{2}{\cT(\trace_A
|\psi_i\rangle\langle\psi_i|)},
$$
where, in order to simplify notations, we have introduced the
shorthand $\cT(\rho):=\rho-\trace[\rho]\id/d$ for the traceless part
of a (non-normalised) $d$-dimensional state.

Since $D$ is invariant under local unitaries, we can take $\psi$ in
Schmidt-diagonal form and put
$$
\psi=(\cos\alpha,0,0,\sin\alpha)^T,
$$
with $0\le \alpha\le \pi/4$.
As $\phi$ is orthogonal to $\psi$, it must be of the form
$$
\phi=(\sqrt{r}\sin\alpha,\sqrt{1-r}\,x,\sqrt{1-r}\,y,-\sqrt{r}\cos\alpha)^T,
$$
with $0\le r\le 1$, $x=\sin\beta$, and $y=\cos\beta e^{i\eta}$, for
$0\le\beta\le\pi/2$. One checks that
$ \|\trace_A\rho-\id/2 \|_2\ge  \|\trace_B\rho-\id/2 \|_2$ if and only if
$|x|\le|y|$, or $\sin\beta\le\cos\beta$, or $0\le\beta\le\pi/4$.

We will show below that, if $0\le\beta\le\pi/4$, then there exists a
$\theta $ and $\chi$ such that
$\cT(\trace_B |\psi_1\rangle\langle\psi_1|) $
$\!= s \cT(\trace_B |\psi_2\rangle\langle\psi_2|)$, for some $s\ge0$.
This implies that, for those $\theta,\chi$,
\beas
\lefteqn{\sum_{i=1,2} \schatten{2}{\cT(\trace_B
|\psi_i\rangle\langle\psi_i|)}} \\
&=& \schatten{2}{\sum_{i=1,2}\cT(\trace_B |\psi_i\rangle\langle\psi_i|)} \\
&=& \schatten{2}{\cT(\trace_B\rho)},
\eeas
so that $D(\rho)\le\sqrt{2}\schatten{2}{\cT(\trace_B\rho)}$.
If, on the other hand, $|x|\ge |y|$, then the same equality can be
made to satisfy with $\trace_B$ replaced by $\trace_A$,
yielding $D(\rho)\le\sqrt{2}\schatten{2}{\cT(\trace_A\rho)}$.

Now define
\beas
K&=& p \,\,\cT(\trace_B |\psi\rangle\langle\psi|) \\
G&=& (1-p)\,\, \cT(\trace_B |\phi\rangle\langle\phi|) \\
H&=& \sqrt{p(1-p)}\,\, (e^{i\chi}\trace_B(|\psi\rangle\langle\phi|)
+ \mbox{h.c.})
\eeas
then
\beas
\cT(\trace_B |\psi_1\rangle\langle\psi_1|) &=& \cos^2\theta\, K
+\sin^2\theta\, G -\sin\theta\cos\theta \,H \\
\cT(\trace_B |\psi_2\rangle\langle\psi_2|) &=& \sin^2\theta\, K
+\cos^2\theta\, G +\sin\theta\cos\theta \,H.
\eeas
We have to find $\theta,\chi$ such that
$\cos^2\theta \,K +\sin^2\theta\, G -\sin\theta\cos\theta \,H $
$\!= s(\sin^2\theta\, K +\cos^2\theta\, G +\sin\theta\cos\theta \,H)$
for some $s\ge0$.
Putting $(s-1)/(s+1)=\cos\gamma$, this is equivalent to finding
$\theta,\chi,\gamma$ such that
\be
\cos(2\theta)(K-G) = \cos\gamma \, (K+G) + \sin(2\theta) H.
\ee
Now $K$, $G$ and $H$ are $2\times 2$ traceless Hermitian matrices, and
are determined by three real parameters: the entries of their
respective Bloch vectors. The previous equation can thus be put in
vector form. In order for it to have a solution for $\theta$, the
three Bloch vectors of $K$, $G$ and $H$ must be linearly dependent,
with real proportionality constants.

\begin{widetext}
Taking into account the special forms of $\psi$ and $\phi$, we have
\beas
K &=& p\twomat{\cos^2\alpha -1/2}{0}{0}{\sin^2\alpha-1/2}, \\
G &=& (1-p) \twomat{r\sin^2\alpha
+(1-r)|x|^2-1/2}{\sqrt{r(1-r)}(\sin\alpha \overline{y}-\cos\alpha x)}
{\sqrt{r(1-r)}(\sin\alpha y-\cos\alpha
\overline{x})}{r\cos^2\alpha+(1-r)|y|^2-1/2}, \\
H &=&
\sqrt{p(1-p)}\twomat{\sqrt{r}\sin(2\alpha)\cos\chi}{\sqrt{1-r}(\cos\alpha
e^{i\chi}\overline{y} + \sin\alpha e^{-i\chi}x)}
{\sqrt{1-r}(\cos\alpha e^{-i\chi}y + \sin\alpha
e^{i\chi}\overline{x})}{-\sqrt{r}\sin(2\alpha)\cos\chi}.
\eeas
\end{widetext}

Expressing the conditions for linear dependence is the first step to
solving our problem. Since the only non-zero component of $K$ is its
$z$-component, the three Bloch vectors are linearly dependent with
real proportionality constants if and only if the (1,2)-entries of $G$
and $H$ (corresponding to their $x$ and $y$ Bloch vector entries) have
the same argument (modulo $\pi$). That is:
$$
\angle(\sin\alpha \overline{y}-\cos\alpha x) = \angle(\cos\alpha
e^{i\chi}\overline{y} + \sin\alpha e^{-i\chi}x)\pmod\pi.
$$
The only unknown here is $\chi$, and it turns out that there always is
a solution. Therefore, the condition of linear dependence fixes $\chi$.
To wit, the solution is
\beas
e^{i\chi} &=& (\sin(2\alpha)-\sin(2\beta)\cos\eta -i
\sin(2\beta)\sin\eta)/Q, \\
Q &=& \sqrt{\sin^2(2\alpha)+\sin^2(2\beta)-2\sin(2\alpha)\sin(2\beta)\cos\eta}.
\eeas

With this choice of $\chi$, we thus have a solution to the equation
$b(K-G) = (K+G) + a H$ in $a$ and $b$. Now we must make sure that this
solution satisfies $a=\sin(2\theta)/\cos\gamma,
b=\cos(2\theta)/\cos\gamma$ for some $\gamma,\theta$. This is so
provided $a^2+b^2\ge 1$. Considering only the $(1,1)$-entries and the
imaginary part of the $(1,2)$-entries of $K,G,H$, and inserting the
solution of $\chi$, we get a $2\times 2$ system of equalities
\beas
-H_{11}a+(K_{11}-G_{11})b &=& K_{11}+G_{11}, \\
-\Im H_{12}a-\Im G_{12}b&=&\Im G_{12},
\eeas
where
\beas
 K_{11} &=& p(\cos^2\alpha-1/2), \\
 G_{11} &=& (1-p)(r\sin^2\alpha+(1-r)\sin^2\beta-1/2),\\
 H_{11} &=& \sqrt{p(1-p)}
 \sqrt{r}\sin(2\alpha)\frac{\sin(2\beta)\cos\eta-\sin(2\alpha)}{Q},\\
 \Im G_{12} &=& -(1-p)\sqrt{r}\sin\alpha\cos\beta\sin\eta\sqrt{1-r},\\
 \Im H_{12} &=&
 \sqrt{p(1-p)}\frac{\cos(2\alpha)
 +\cos(2\beta)}{Q}\cos\beta\sin\eta\sqrt{1-r}.
\eeas
We have been able to show that the solution $(a,b)$ of this system,
with these rather formidable expressions for the coefficients, indeed
satisfies $a^2+b^2\le1$ for $0\le \beta\le \pi/4$, and $a^2+b^2\ge 1$
for $\pi/4\le\beta\le\pi/2$. This was done using a computer algebra
system, and we refer the interested reader to the supplementary
material \cite{suppl}. This proves the Proposition for rank 2 states.
\qed

%%%%%%%%%%%%%%%%%%%%%%%%%%%%%%%%%%%%%%%%%%%%%%%%%%%%%%%%%%%%%%%%%%%%%
\section{Entanglement of Assistance}\label{sec:eoa}

In this final Section, we briefly touch upon another approach to
characterise \G\ maps which is still similar in spirit to the one
discussed before. Rather than using quadratic relations to characterise
whether a pure state is maximally entangled, one can look at the
entropy of entanglement of the state. Applying a similar procedure as
above then yields the so-called entanglement of assistance. The
entanglement of assistance, $E_A$, of a state \cite{assistance} is in
some sense the converse of the entanglement of formation, $E_F$. It is
defined as
\be
E_A(\rho) = \max_{\{p_i,\psi_i\}} \big\{\sum_i p_i E(\psi_i): \sum_i
p_i \ket{\psi_i}\bra{\psi_i} = \rho\big\},
\ee
whereas replacing the maximisation by a minimisation yields the
entanglement of formation. In other words, while $E_F$ is the convex
hull of the pure state entanglement functional $E$, $E_A$ is its
concave hull.

The connection between this quantity and the problem of determining
\G\-ness of a map is quite clear: a $d$-dimensional CP map is a \G\
map if and only if its state representative is a convex combination of
maximally entangled pure states, if and only if that state
representative has the maximal possible $E_A$ of $\log d$.

Fo qubit maps one has the alternative of using the concurrence of
assistance, $C_A$, defined in \cite{concass} as the concave hull of
the pure state concurrence. Thus, a $2$-dimensional CP map is a \G\
map if and only if its state representative has the maximal possible
$C_A$ of $2$. A closed-form expression for $C_A$ is given by
\cite{concass}
\be
C_A(\rho) =  \|\rho^{1/2}(\sigma_y\otimes\sigma_y)\rho^{1/2} \|_1.
\ee

%%%%%%%%%%%%%%%%%%%%%%%%%%%%%%%%%%%%%%%%%%%%%%%%%%%%%%%%%%%%%%%%%%%%%%
\section{Conclusions}

The set of completely positive maps can be distinguished by their
decomposability into a convex combination of unitary conjugations. We have
shown that one can find necessary and sufficient conditions for
determining whether a given CPT map belongs to the set of \G\ maps.
The criterion requires the collective ``off-diagonalization'' of
$d^2-1$ matrices of dimension $d^2\times d^2$ that are built from the
$d^2-1$ basis elements of the Hilbert space into which the density matrices
on $\cH$ are embedded.

Based on this criterion, we have defined a proper distance measure to the
set of \G\ matrices which can be used to quantify the extent to
which non-classical error mechanisms have influenced the evolution of a
quantum system. In this way, it would be possible to point towards the
dominating error source in a specific physical realization of the CPT map.
This means that tomographic reconstructions of physical processes can
provide a host of valuable information about the process itself, of which
the amount of non-classical error mechanisms is but one. We believe that
a thorough investigation of tomographic process reconstruction can reveal
a plethora of information about the underlying physical mechanisms that
led to the realization of the CPT map under investigation, and which has
hitherto not been fully appreciated.

%%%%%%%%%%%%%%%%%%%%%%%%%%%%%%%%%%%%%%%%%%%%%%%%%%%%%%%%%%%%%%%%%%%%%%
\acknowledgments
SS thanks the UK Engineering and Physical Sciences Research
Council (EPSRC) for support. The work of KMRA is part of the QIP-IRC
(\texttt{www.qipirc.org}), supported by EPSRC (GR/S82176/0). KMRA is
also supported by the Institute of Mathematical Sciences, Imperial
College London.

%%%%%%%%%%%%%%%%%%%%%%%%%%%%%%%%%%%%%%%%%%%%%%%%%%%%%%%%%%%%%%%%%%%%%%

\end{document}